\documentclass[preprint2]{aastex}
\usepackage{amsmath} 
\usepackage{bm}

\shorttitle{Reconnection-Driven Turbulence in a Coronal-Hole Jet}
\shortauthors{Uritsky et al.} 

\begin{document}

\title{Reconnection-Driven Magnetohydrodynamic Turbulence \\
in a Simulated Coronal-Hole Jet}

\author{Vadim M.\ Uritsky$^{1,2}$, Merrill A.\ Roberts$^{1,2}$, C.\ Richard DeVore$^2$, and Judith T.\ Karpen$^2$ }

\affil{$^1$Catholic University of America, 620 Michigan Avenue NE, Washington, DC 20064 USA \\
$^2$Heliophysics Science Division, NASA Goddard Space Flight Center, Greenbelt, MD 20771 USA}

\email{vadim.uritsky@nasa.gov}






\begin{abstract}

Extreme-ultraviolet and X-ray jets occur frequently in magnetically open coronal holes on the Sun, especially at high solar latitudes. Some of these jets are observed by white-light coronagraphs as they propagate through the outer corona toward the inner heliosphere, and it has been proposed that they give rise to microstreams and torsional Alfv\'{e}n waves detected {\em in situ} in the solar wind. To predict and understand the signatures of coronal-hole jets, we have performed a detailed statistical analysis of such a jet simulated with an adaptively refined magnetohydrodynamics model. The results confirm the generation and persistence of three-dimensional, reconnection-driven magnetic turbulence in the simulation. We calculate the spatial correlations of magnetic fluctuations within the jet and find that they agree best with the M\"{u}ller - Biskamp scaling model including intermittent current sheets of various sizes coupled via hydrodynamic turbulent cascade. The anisotropy of the magnetic fluctuations and the spatial orientation of the current sheets are consistent with an ensemble of nonlinear Alfv\'{e}n waves. These properties also reflect the overall collimated jet structure imposed by the geometry of the reconnecting magnetic field. A comparison with Ulysses observations shows that turbulence in the jet wake is in quantitative agreement with that in the fast solar wind. 

\end{abstract}

\keywords{Magnetic reconnection -- Sun: corona -- Sun: magnetic fields -- Turbulence -- Waves}


\section{Introduction}
\label{sec:intro}


Magnetohydrodynamic (MHD) turbulence plays a fundamental role in numerous systems of critical interest to heliophysics: the solar convective zone and atmosphere; the interplanetary medium; the magnetotails of the Earth and other magnetized planets; and the heliopause, at the interface between the heliosphere and the interstellar medium. In plasmas characterized by moderate to high values of plasma $\beta \equiv n k_B T /(B^2 / 2 \mu_0)$,
where $n$ and $T$ are plasma number density and temperature, $B$ is magnetic field strength, $k_B$ is Boltzmann's constant, and $\mu_0$ is the permeability of free space,  thermal pressure reaches or exceeds the pressure exerted by the magnetic field, and the turbulent flows readily bend and fold the field to produce current sheets throughout the volume. Magnetic reconnection or resistive diffusion across these sheets converts magnetic free energy to kinetic and thermal energies of the bulk plasma and to kinetic energy of highly accelerated particles. In low-$\beta$ plasmas, where the magnetic field is dominant, the current sheets form and the associated reconnection/diffusion processes occur preferentially near null points of the field. Null regions therefore act as generators of reconnection-driven turbulence in highly conducting, low-$\beta$ plasmas such as the solar corona.

In general, turbulence occurs naturally in moving fluids characterized by a high Reynolds number, $Re = v_l l / \nu$, where $v_l$ is a typical ambient flow speed at the scale $l$ and $\nu$ is the kinematic viscosity. (In resistive MHD plasmas, the magnetic Reynolds or Lunquist number is defined using the magnetic diffusivity $\eta$ in place of the viscosity.) The condition $Re \gg 1$, satisfied in many astrophysical systems, ensures that a broad range of inertial spatial scales $l$, bounded by the large driving scale $l_D$ and the small dissipative scale $l_d$, $l_d \ll l \ll l_D$, is supported. While the physical mechanisms underlying energy dissipation at the dissipative scale $l_d$ play a critical role by providing a sink for the energy injected into the system at the driving scale $l_D$, the inertial-range behavior in between is, to a large extent, independent of the details of the dissipation process. This leads to a statistical self-similarity of fluctuations in the velocity and magnetic fields across those scales, even in weakly collisional to collisionless plasmas where the dissipation is dominated by wave-particle interactions or other kinetic effects \citep{schekochihin2009,daughton2011,leonardis13}.

The turbulent outflows from reconnection-driven systems such as coronal jets can provide important clues about the geometry of the reconnection region and enable remote sensing of the driving mechanism through its characteristic signatures in the flow. Such features include unstable velocity shear and ensuing multiscale vorticity in the reconnection exhaust of terrestrial substorms \citep{keiling09}, fragmented current sheets and filaments embedded in larger-scale outflow from field-reversed configurations \citep{uritsky01, klimas04}, and topological markers of the underlying magnetic-field configuration through the three-dimensional (3D) geometry of the turbulent flows \citep[e.g.\ ][]{biskamp03}. These and other observational hallmarks of the reconnection process have been identified in extreme ultraviolet (EUV) images of the corona from {\em Solar and Heliospheric Observatory} and {\em Solar Terrestrial Relations Observatory } spacecraft \citep{uritsky07,  uritsky13}, flyby time-series data from {\em MErcury Surface, Space ENvironment, GEochemistry, and Ranging} probe \citep{uritsky11}, and numerical data from high-resolution 3D simulations of prescribed, vorticity-laden MHD flows \citep{uritsky10a}.

Solar EUV and X-ray jets \citep[][and references therein]{raouafi2016} are transient, highly dynamic brightenings of the low solar corona that produce fast collimated outflows of plasma. When these events occur in the open magnetic fields of coronal holes, the quasi-radial jet outflows sometimes are observed by white-light coronagraphs to propagate several solar radii from the Sun into the inner heliosphere. \citet{neugebauer2012} suggested that these jets may be the origin of small-scale microstreams detected {\em in situ} in the solar wind \citep{neugebauer1995}. Because many jets sensed remotely in the corona exhibit a distinctively helical structure that traverses the corona at highly supersonic speeds, leading to its identification as a nonlinear Alfv\'en wave, it is plausible that coronal hole jets also are the source of such waves detected in the interplanetary medium \citep{gosling10,marubashi10}.

In this paper, we analyze a first-principles, 3D numerical simulation of the initiation and propagation of a coronal hole jet \citep{karpen16} to establish and characterize its turbulent nature. The physical model underlying the simulation is null-point magnetic reconnection occurring at the interface between the ambient coronal-hole flux of one polarity and a concentrated patch of opposite-polarity flux provided by an embedded bipole \citep{lau1990,antiochos1990}. The topology of this configuration, in which an inner system of flux that closes to the solar surface is embedded within an outer system of flux that opens to the heliosphere, supports strong electric currents associated with steep gradients in the magnetic-field direction at the interface between the two flux systems \citep{antiochos1996}. Previous Cartesian, gravity-free, uniform-background 3D numerical simulations of such configurations \citep{pariat09, pariat10, pariat15, pariat16} demonstrated that this model produces explosive jets with helical structure, density-enhanced outflows, and Alfv\'enic wave fronts, in accord with observations. The energy source for the jets is the twisted magnetic flux under the separatrix. The new work \citep{karpen16} extends those investigations by including the effects of spherical geometry, solar gravity, density and magnetic-field stratification, and an isothermal solar wind. The ensuing reconnection-driven jet wave front propagates unhindered into the outer corona, reaching 5 solar radii in 1250 s, and its duration, length, diameter, plasma-outflow speed, leading-front speed, plane-of-the-sky transverse speed, kinetic energy, and helical morphology are all typical of observed coronal hole jets.

We review the relevant statistical hierarchical models of turbulent hydrodynamic and magnetohydrodynamic cascades, with and without intermittency, in \S \ref{sec:turb}. A concise summary of the numerical simulation, which is described in detail by \citet{karpen16}, is given in \S \ref{sec:model}. The main results of our analysis of the jet are presented in the succeeding three sections. We discuss noteworthy large-scale features of its radial structure in \S \ref{sec:spat}, analyze the statistics of its small-scale velocity- and magnetic-field fluctuations in \S \ref{sec:stat}, and illustrate its filamentary electric-current structures in \S \ref{sec:cs}. The paper concludes with a brief discussion of the implications of our results in \S \ref{sec:disc_conc}.

\section{Hierarchical MHD Turbulence Models}
\label{sec:turb}

%
A statistical framework for describing inertial-range turbulence was introduced by \citet[][hereafter referred to as the K41 model]{kolmogorov41} for hydrodynamic fluids. He assumed that the energy dissipation is spatially uniform and isotropic, and that cross-scale interactions take place through a local cascade process constantly breaking turbulent eddies into smaller pieces. Consequently the (uniform) energy dissipation rate $\epsilon$ per unit mass scales as follows:
\begin{equation}
\epsilon \propto \left( \delta v_l \right)^2 / \tau_l \propto \left( \delta v_l \right)^3 / l.
\label{eq_eps}
\end{equation}
Here $\delta v_l$ is the characteristic velocity magnitude and $\tau_l$ is the typical eddy turnover time at scale $l$. This relation leads directly to the primary result of the K41 theory:
\begin{equation}
\delta v_l \propto \epsilon^{1/3} l^{1/3}.
\label{eq_delv}
\end{equation}
An important consequence of the $l^{1/3}$ scaling of the turbulent velocity field is that the kinetic energy content $E_k$, obtained by integrating the kinetic energy over all wavenumbers beyond $k$, obeys the familiar law
\begin{equation}
E_k \propto \epsilon^{2/3} k^{-\alpha}
\label{eq_espec}
\end{equation}
with spectral index $\alpha = 5/3$. 

The K41 scaling relations have proved highly successful in analyses of turbulent energy spectra \citep[e.g.][and references therein]{bruno13}. However, typically they fail to explain the higher-order statistics of turbulent flows. The K41 theory assumes that $\epsilon^p$ is independent of $l$ for all orders $p$, with the result that 
\begin{equation}
\delta v_l^p \propto l^{\zeta_p}
\label{eq_delvp}
\end{equation}
and
\begin{equation}
\zeta_p = p/3.
\label{eq_K41}
\end{equation}
The ${\zeta_p}$ are known as the structure function exponents. The first-order exponent replicates the velocity scaling in Equation (\ref{eq_delv}), $\zeta_1 = 1/3$, while the second-order exponent relates to the energy spectral index in Equation (\ref{eq_espec}), $\alpha = \zeta_2 + 1 = 5/3$. 

In practice, however, it is often found that the linear approximation in Equation (\ref{eq_K41}) is invalid. Turbulent intermittency causes the energy dissipation rate $\epsilon_l$ to vary with spatial scale $l$.

Intermittency in turbulent flows reflects the presence of intense small-scale dissipative structures breaking global scale-invariance \citep{she94, boratav97, uritsky07, abramenko10}. Such structures are commonly observed both in laboratory experiments and in nature, and are successfully reproduced in high-resolution numerical simulations \citep[for a review, see][]{she09}. To account for these effects, the energy dissipation rate is assumed to follow the Kolmogorov refined similarity (KRS) hypothesis \citep{kolmogorov62, stolovitzky94}, 
\begin{equation}
\epsilon_l^p \propto l^{\tau_p}.
\label{eq_epsp}
\end{equation}
The new exponents $\tau_p$ are related to $\zeta_p$ via 
\begin{equation}
\zeta_p = p/3 + \tau_{p/3}.
\label{eq_KRS}
\end{equation}
The particular values 
\begin{align}
\zeta_0 &= 0~~~\tau_0 = 0,\nonumber \\
\zeta_3 &= 1~~~\tau_1 = 0,
\label{eq_KRS_exact}
\end{align}
are exact results of the conservation of energy in steady incompressible fluids under the assumptions of isotropy and homogeneity \citep{kolmogorov62,  politano95}. These values also are approximately correct for a much wider class of flows in which the local dissipation rates and the velocity fluctuations exhibit the strong correlation underlying the KRS scaling.

The key step in deriving the appropriate statistical description of intermittent turbulence is to determine the $p$ dependence of $\tau_p$ (i.e., the departure of $\zeta_p$ from the $p/3$ law) for $p \ge 2$. For inertial-range hydrodynamic turbulence that is fully described in terms of moment ratios of the energy dissipation rate $\epsilon_l$, \citet[][hereafter the SL model]{she94} obtained the closed-form solution
\begin{equation}
\tau_{p/3} = -2p/9 + 2[1-(2/3)^{p/3}].
\label{eq_tau_SL}
\end{equation}
This immediately yields, from Equation (\ref{eq_KRS}),  
\begin{equation}
\zeta_p = p/9 + 2[1-(2/3)^{p/3}]. 
\label{eq_SL}
\end{equation}
The SL model satisfies Equations (\ref{eq_KRS_exact}) and predicts the energy spectrum exponent $\alpha \approx 1.696$, within $2\%$ of the K41 prediction in Equation (\ref{eq_K41}). However, SL deviates strongly from K41 for higher statistical moments affected by intermittent energy dissipation in quasi-one-dimensional vortex filaments. The scaling corrections introduced by SL theory have been verified in many experimental and theoretical studies \citep{she09}. The degree of intermittency is evaluated using higher order ($p > 3$) structure functions requiring extensive spatial averaging \citep{abramenko08}. 

By comparison with hydrodynamic turbulence, magnetohydrodynamic turbulence exhibits a rather different form of scaling, influenced by the coupling of magnetic and velocity perturbations by Alfv\'{e}n waves. These effects in non-intermittent (uniform $\epsilon_l$) magnetofluid turbulence were first investigated independently by \citet{iroshnikov63} and \citet{kraichnan65}; hereafter we refer to them jointly as the IK model. This coupling is usually described in terms of Els{\"a}sser variables $\bm{z^{\mp} = v \pm B}$. In the IK framework, $\bm{z^+}$ and $\bm{z^-}$ eddies interact when they meet while traveling in opposite directions along the large-scale magnetic field $\bm{B_0}$. Relative to kinematic eddies in hydrodynamic turbulence, Alfv\'{e}nic eddies in the Els{\"a}sser field have a cross-scale energy transfer rate reduced by the factor $\tau_A/\tau_l$, where $\tau_A = l/v_A$ is the Alfv\'{e}n time at scale $l$ and $v_A$ is the Alfv\'{e}n speed. Due to this weak interaction, 
\begin{equation}
\epsilon \propto \left( \tau_A / \tau_l \right) \left( \delta z_l \right)^2 / \tau_l \propto \tau_A \left( \delta z_l \right)^4 / l
\label{eq_IKeps}
\end{equation}
so that the Els{\"a}sser perturbations scale as
\begin{equation}
\delta z_l \propto (\epsilon v_A)^{1/4} l^{1/4}.
\label{eq_IKdelz}
\end{equation}
In weakly compressible regimes \citep{zank93}, the scaling behavior of velocity and magnetic field perturbations repeats that of $\delta z_l$ \citep{she09}. This leads to the energy spectrum 
\begin{equation}
E_k \propto (\epsilon v_A)^{1/2} k^{-3/2},
\label{eq_IKespec}
\end{equation}
with spectral index $\alpha = 3/2$.

To describe the scale-dependent dissipation regimes that typically occur in magnetized plasmas, the statistical formalism developed by \citet{she94} has been extended to include the IK phenomenology \citep{grauer94,politano95}. The KRS then takes the form 
\begin{equation}
\zeta_p = p/4 + \tau_{p/4}
\label{eq_KRS_MHD}
\end{equation}
with 
\begin{align}
\zeta_0 &= 0~~~\tau_0 = 0,\nonumber \\
\zeta_4 &= 1~~~\tau_1 = 0,
\label{eq_KRS_MHD_exact}
\end{align}
in analogy with Equations (\ref{eq_KRS_exact}) from hydrodynamics. The dissipation rate exponents describing the intermittent turbulence follow the rule 
\begin{equation}
\tau_{p/4} = -p/8 + [1 - (1/2)^{p/4}].
\label{eq_tau_PP}
\end{equation}
This results in \citep[][hereafter PP]{politano95}
\begin{equation}
\zeta_p = p/8 + 1 - (1/2)^{p/4}.
\label{eq_PP}
\end{equation}
The energy spectrum associated with the second-order exponent $\zeta_2$ in the intermittent PP model has $\alpha \approx 1.543$. This correction is slightly larger (about $3\%$) than that of the SL model for non-magnetized fluids, while the higher moments display still more significant departures from the non-intermittent scaling.

Subsequent 3D magnetohydrodynamic turbulence simulations \citep[][hereafter the MB model]{muller00} have exhibited a hybrid behavior between the SL and PP models, in which K41-like scaling supported by a vortex cascade combined with IK-like dissipative structures in the form of current sheets. In this case, the $\zeta_p$ exponents take the form in Equation (\ref{eq_KRS}) with 
\begin{equation}
\tau_{p/3} = -2p/9 + [1-(1/3)^{p/3}].
\label{eq_tau_MB}
\end{equation}
Then
\begin{equation}
\zeta_p = p/9 + 1 - (1/3)^{p/3},
\label{eq_MB}
\end{equation}
and the energy spectrum in the intermittent MB model has $\alpha \approx 1.741$. This model also satisfies Equations (\ref{eq_KRS_exact}).

The three hierarchical intermittency models in Equations (\ref{eq_SL}), (\ref{eq_PP}), and (\ref{eq_MB}), as well as a number of models proposed more recently, are conveniently represented by a unifying scaling ansatz containing three adjustable parameters $q$, $s$, and $d$ \citep{muller00}:
\begin{equation}
\zeta_p =(p/q)(1-s) + d[1 - (1 - s/d)^{p/q}].
\label{eq_hier}
\end{equation}
Here $q$ is a spatial exponent describing the basic scaling $\delta v_l \propto l^{1/q}$, $s$ is a temporal exponent reflecting energy transfer time at the smallest inertial scales, $t_l \propto l^s$, and the codimension $d=3-D$ is defined by the dimension $D$ of dissipative structures embedded in three-dimensional space (Table \ref{tab_models}). The SL hydrodynamic model with vortex filaments ($D = 1$) is recovered by substituting $q = 3$, $s = 2/3$, and $d = 2$ into Equation (\ref{eq_hier}). The PP magnetohydrodynamic model is obtained for $q = 4$, $s = 1/2$, and $d = 1$, with the dissipative structures interpreted as current sheets. The combination $q = 3$, $s = 2/3$, and $d = 1$ yields the MB model. More complex intermittency regimes can be represented using Equation (\ref{eq_hier}), including anisotropic scaling of velocity, magnetic and Els{\"a}sser field perturbations \citep[see][and references therein]{she09}.

\begin{table*}
\begin{center}
\caption{Classification of turbulence models based on Equation (\ref{eq_hier})}
\begin{tabular}{ l c c c c c }
\hline
 Model 												& $q$ & $s$ & $d$ & Cascading structures & Dissipative structures \\
\hline
{\it Non-intermittent} \\
Kolmogorov 1941 (K41)					& 3	& 0 & - & Fluid vortices & - \\
Iroshnikov \& Kraichnan (IK)	& 4	& 0 & - & Alfv\'{e}n wave packets & - \\
Brownian noise (BN)						& 2 & 0 & - & - & - \\
\hline
{\it Intermittent} \\
She \& Leveque (SL)						& 3	& 2/3 & 2 & Fluid vortices & Vortex filaments \\
Politano \& Pouquet (PP)			& 4 & 1/2 & 1 & Alfv\'{e}n wave packets & Current sheets \\
M{\"u}ller \& Biskamp (MB)		& 3 & 2/3 & 1 & Fluid vortices & Current sheets \\
\hline
\end{tabular}
\label{tab_models}
\end{center}
\end{table*}

By way of contrast, the classical Brownian noise model (hereafter the BN model) with uncorrelated fluctuations corresponds to $q = 2$ and $s = 0$ ($d$ arbitrary) in Equation (\ref{eq_hier}), i.e.,
\begin{equation}
\zeta_p = p/2.
\label{eq_BN}
\end{equation}
As will be demonstrated in \S \ref{sec:stat}, our coronal-jet simulations deviate strongly from the uncorrelated BN model, as well as from non-intermittent magnetic turbulence. They adhere most closely to the MB model expressed in Equation (\ref{eq_MB}).


\section{Coronal Jet Simulation}
\label{sec:model}

\begin{figure*}
\begin{center}
\includegraphics[width=12.5 cm]{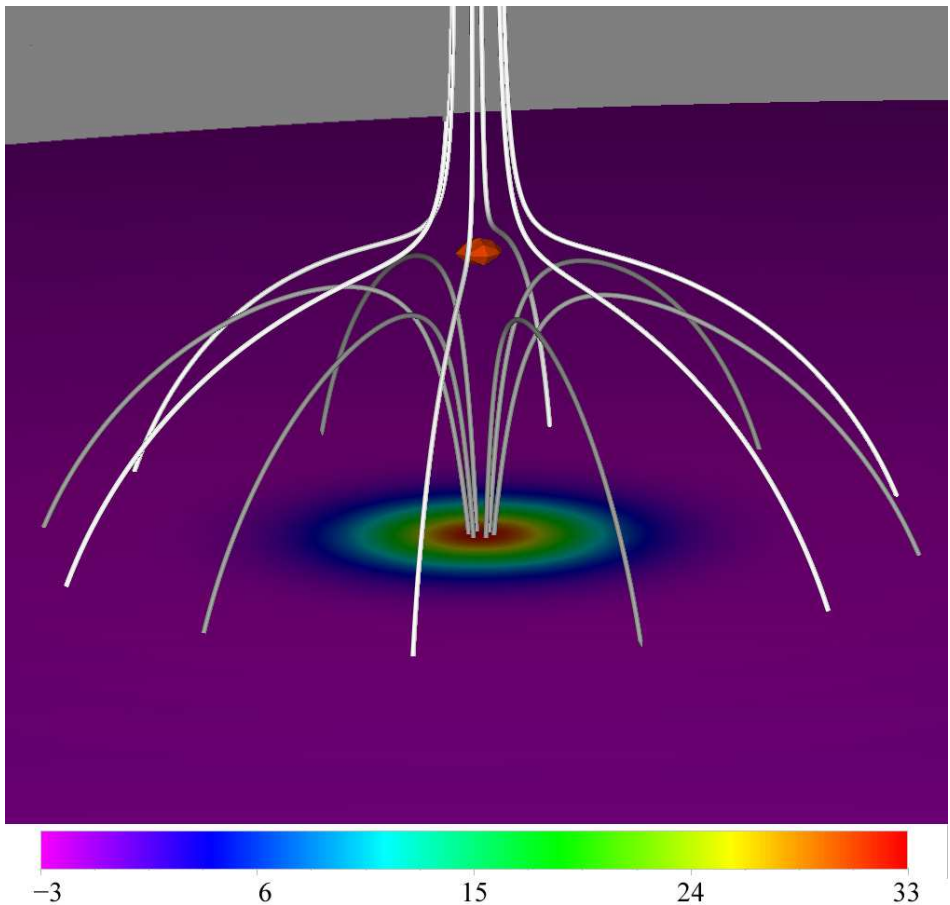}
\caption{\label{fig_BB_zoomin} Close view of jet-producing region at $t = 0$ s, showing magnetic field lines in flux systems that are open (white curves) and closed (gray curves), the sign and magnitude of the radial component of the magnetic field on the solar surface ($10^{-4}$T, color shading), and the high-$\beta$ region encompassing the magnetic null point (suspended spheroid)}.
\end{center}
\end{figure*}

\begin{figure*}
\begin{center}
\includegraphics[width=14. cm]{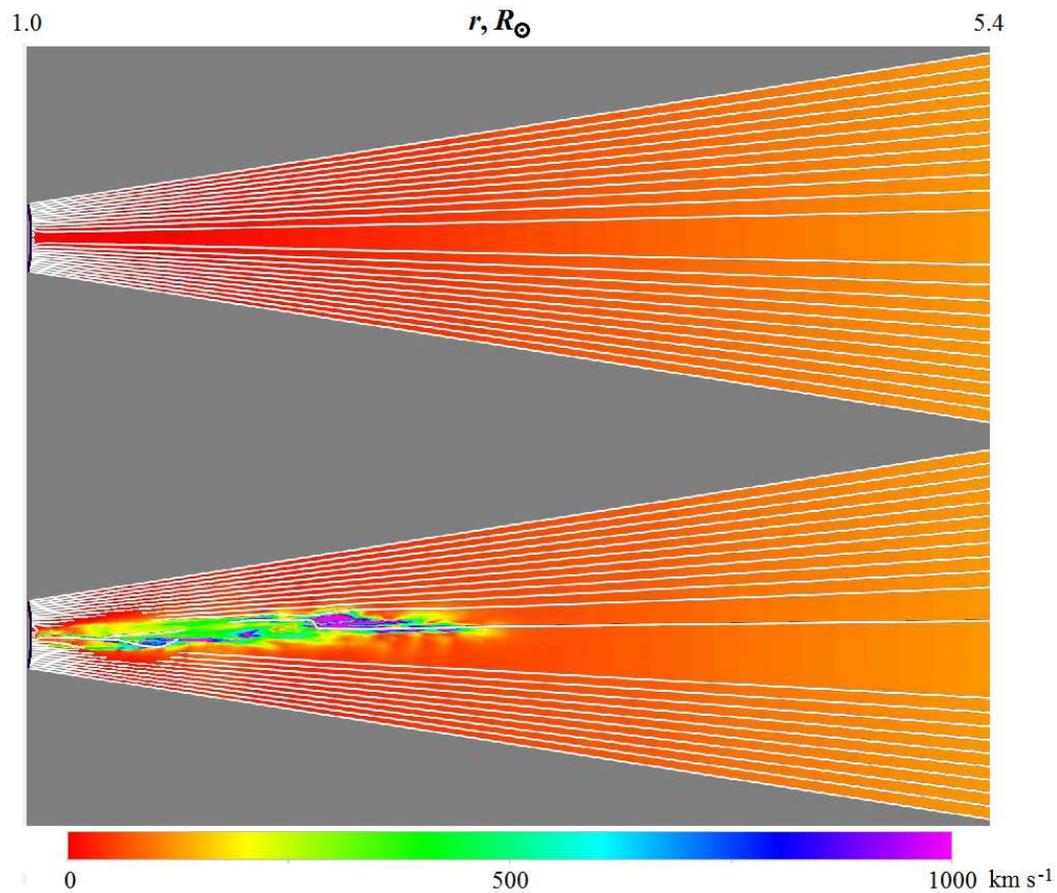} 
\caption{\label{fig_VB_zoomout} Far view of jet cross-section (at $\theta = 0^{\circ}$) showing magnetic field lines (white lines) and velocity magnitude (km s$^{-1}$, color shading) against the plane of the sky at $t = 0$ s (top) and $t = 3650$ s (bottom). An animation of this figure for $t \in [0,4000]$ s at 25 s cadence is provided online.}
\end{center}
\end{figure*}

The coronal-jet numerical simulation that we analyze is described in considerable detail by \citet{karpen16}. Here, we summarize its main features to provide context for the turbulence analysis presented in the following sections. The magnetic structure of the jet-producing region is shown at time $t$ = 0 in Figure \ref{fig_BB_zoomin}, illustrating the inner closed (gray curves) and outer open (white curves) flux systems and the dome-shaped separatrix between them with a magnetic null point (red spheroid) at its top. Two still images and a corresponding online animation from the simulation comprise Figure \ref{fig_VB_zoomout}, which shows the spherical domain to height $r \approx {\bf 5.4}R_\odot$, selected magnetic field lines (white lines), and plasma velocity magnitude against the plane of the sky (color shading) at times $t$ = 0 s (top) and 3650 s (bottom). The jet onset was at $t \approx$ 2750 s, and the simulation was stopped at $t$ = 4000 s.

A spherical domain of extent $[1R_\odot,9R_\odot] \times [-9^\circ,+9^\circ] \times [-9^\circ,+9^\circ]$ in radius, latitude, and longitude was assumed, with open inner and outer radial boundaries and closed side boundaries in both transverse directions. $R_\odot = 7 \times 10^{8}$ m is the radius of the Sun. The grid was exponentially stretched radially and linearly spaced in angle to divide the domain into nearly cubic cells throughout. Adaptive mesh refinement, managed by the PARAMESH toolkit optimized for parallel computer architectures \citep{macneice2000}, was used to target local regions where the electric current density was relatively strong. The volume of maximally refined grid blocks expanded with time to resolve the surface currents that built up at the interface between the open and closed flux systems and, after jet initiation, the filamentary volume currents associated with the propagating nonlinear Alfv\'en waves generated by the onset of impulsive reconnection. The combination of refinements increased the total number of grid cells by an order of magnitude during the simulation, which enabled us to resolve small intermittent structures.

The time-dependent, ideal, magnetohydrodynamic equations for conservation of mass, momentum, and magnetic flux were solved in spherical coordinates using the Adaptively Refined MHD Solver \citep[ARMS;][]{devore2008}. The MHD energy equation was eliminated by assuming that the plasma evolution is isothermal, which is the simplest and most computationally efficient way to generate a supersonic solar wind \citep{parker1958}. ARMS employs Flux-Corrected Transport methods \citep{devore1991} to advance the resulting set of finite-volume equations on the adapted grid. Theoretical developments in implicit large-eddy simulation techniques \citep{grinstein2007} have demonstrated that certain numerical algorithms, when applied to ideal hydrodynamic systems, incorporate an implicit sub-grid-scale model that emulates the viscous diffusion explicitly included in the Navier-Stokes equations. Such schemes are particularly well-suited for simulating large-$Re$ turbulent flows. Flux-Corrected Transport is a member of this class of algorithms \citep{drikakis2007}.

Parameter values chosen for the simulation are representative of coronal hole jets observed on the Sun. The base mass density $\rho_\odot = 2.0 \times 10^{-13}$ kg m$^{-3}$ and uniform temperature $T_\odot = 1.0 \times 10^6$ K are typical of the tenuous, relatively cool atmosphere of coronal holes. Combined with solar gravity, the resulting thermal pressure causes the outflowing wind to become transonic at $r \approx 6R_\odot$ where it reaches $130$ km s$^{-1}$. A locally uniform, ambient coronal-hole field was provided by a Sun-centered monopole of strength $B_m = -2.5 \times 10^{-4}$ T at the surface. The embedded polarity of the jet-generating region was provided by a radially oriented dipole of surface strength $B_d = +3.5 \times 10^{-3}$ T, placed at $(0^\circ,0^\circ)$ in latitude and longitude and at depth $10^7$ m below the surface. As shown in Figure \ref{fig_BB_zoomin}, the superposition of these two fields has a magnetic null point at height $1.5 \times 10^7$ m above the surface. The footprint of the dome-shaped separatrix on the surface is a circle of radius $2.2 \times 10^7$ m, which is typical of jet-generating regions on the Sun.

The magnetic field within the dome is slowly energized by subsonic rotational motions imposed at the base to mimic the photospheric motions. The peak speed $v_{max} = 25$ km s$^{-1}$, about 20\% of the sound speed and less than 1\% of the local Alfv\'en speed. These flows were ramped up smoothly from zero at $t$ = 0 s to full speed at $t$ = 1000 s, then held fixed thereafter. By time $t \approx$ 2750 s, when the reconnection started and began to generate the jet, the motions had introduced about 1.5 turns of maximum twist into the magnetic field beneath the dome. The reconnection onset was driven by an ideal kink-like instability of the twisted magnetic field beneath the dome, as established by preceding Cartesian simulations \citep{pariat09,rachmeler2010}. The resulting buckling of the dome drives together internal twisted field and external untwisted field across the separatrix electric current sheet, transferring twist onto the external field lines where it can propagate away freely into the heliosphere in the form of nonlinear Alfv\'en waves. Wave pressure compresses the plasma, enhancing the density of the jet material trailing the Alfv\'enic wave front above the local, ambient solar-wind value. The front progresses at the coronal Alfv\'en speed (up to $3000$ km s$^{-1}$) in the frame moving with the solar wind, so it can traverse one solar radius in less than 250 s, as indicated by Figure \ref{fig_VB_zoomout}.

\section{Grid Regularization}

Although the adaptive grid utilized by ARMS is essential to resolving important jet features in the model while maintaining manageably sized data files, it presents a formidable challenge when it comes to analyzing those same features.  In order to examine the properties of the jets, especially the statistical self-similarity of the velocity and magnetic fields, a regular grid is necessary. We therefore developed a method of regularizing the grid while maintaining the integrity of the irregularly gridded data. Due to its simplicity and bias towards nearest neighbor points in determining the values of the new grid points, a linear interpolation based on a Delaunay triangulation of the irregular points \citep{delaunay34} was chosen. 

The regularization was implemented in the Interactive Data Language (IDL) using the \textit{Triangulate} and \textit{Trigrid} functions as follows.  After selecting a given two-dimensional slice of the model space with $\phi=0$, the irregularly gridded $r$ and $\theta$ coordinates were read into IDL.  From this set of co-planar points, the \textit{Triangulate} procedure constructs a Delaunay triangulation, using the ``divide and conquer method'' described in \citet{lee80}. The Delaunay triangulation produces a set of triangles from a set of irregular points, such that a circle formed by connecting the vertices of any triangle does not contain any other point.  This maximizes the minimum angle of the triangles, and thus ensures that only nearby points are used to construct them, making it an optimal choice for the correct interpolation and display of our irregularly-gridded data. After the triangles have been created, a spatially regular grid is overlaid on them by the \textit{Trigrid} function, and the value at each point of the regular grid is given by a linear interpolation of the values of the vertices of the Delaunay triangle into which it falls. This again ensures that, since the circumcircle of any Delaunay triangle contains no other points, the values used in the interpolation are those closest to the desired regular grid point, thereby minimizing the distortion of the original dataset.


\section{Radial Structure of the Jet}
\label{sec:spat} 

The structure of the jet is substantially non-uniform at small and intermediate spatial scales, and exhibits a large-scale variability in the radial direction. In this section, we investigate both of these effects at $t=3650$ s, providing a characteristic example of a fully developed turbulent cascade in the model. Temporal evolution of the jet fluctuations will be addressed in the subsequent section. 

\begin{figure*}
\begin{center}
\includegraphics[width=14. cm]{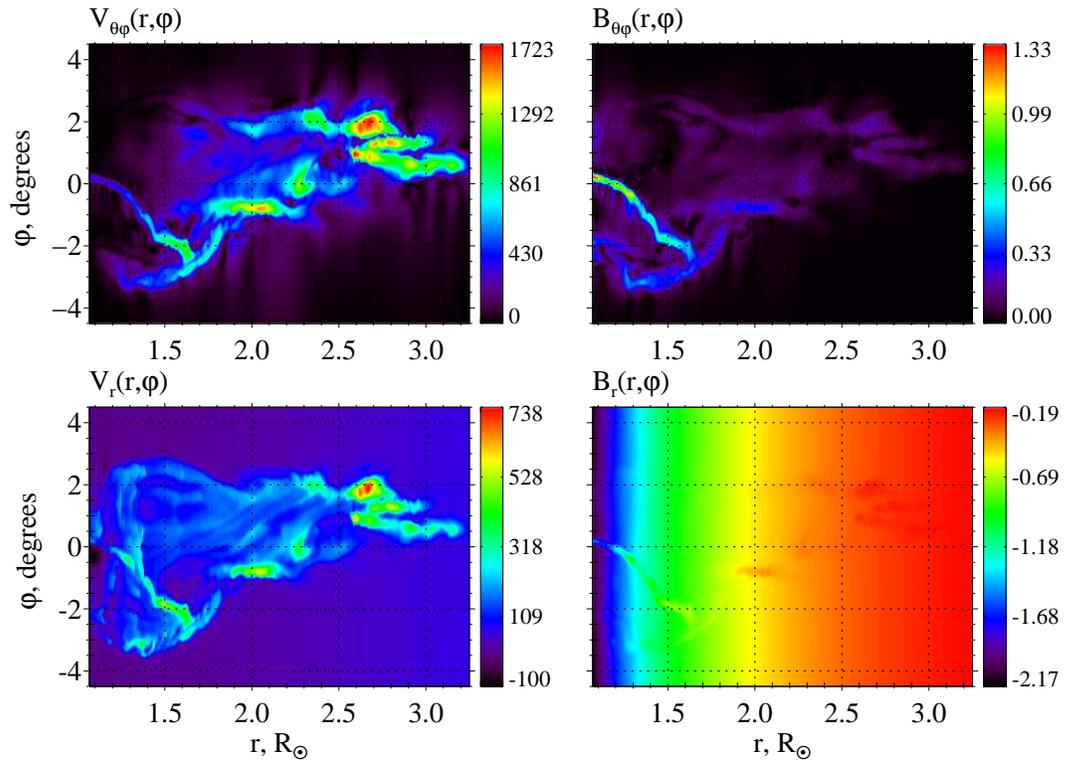}
\caption{\label{fig_VB_img} Constant-latitude cross-sections (at $\theta = 0^{\circ}$) of transverse (top) and radial (bottom) components of velocity (left) and magnetic (right) fields at $t = 3650$ s.}
\end{center}
\end{figure*}

Figure \ref{fig_VB_img} shows constant-latitude cross-sections of the transverse and radial components of the velocity and magnetic fields at $t = 3650$ s as an example of fully developed turbulence. The transverse components are defined as 
\begin{align}
v_{\theta\phi} &\equiv \left( v_\theta^2 + v_\phi^2 \right)^{1/2},\nonumber \\
B_{\theta\phi} &\equiv \left( B_\theta^2 + B_\phi^2 \right)^{1/2}.
\label{deltavb}
\end{align}

The range of $r$ coordinates plotted here excludes the strongest magnetic field near the surface, for better visibility of less intense structures forming at higher altitudes. The velocity field inside the jet volume varies from less than 200 to more than 1700 km s$^{-1}$, exhibiting compact regions of high-speed perturbations embedded in a slower background flow. The magnetic field perturbation also varies by a significant amount. The transverse components of the velocity and the magnetic field have a similar spatial structure spanning across multiple scales as expected for a turbulent flow. A closer investigation of the plots shows that the resemblance between the spatial patterns of $v_{\theta\phi}$ and $B_{\theta\phi}$ increases with $r$. As demonstrated below, this reflects the dominance of Alfv\'{e}nic coupling at higher altitudes. The correlation between the radial field components plotted in the bottom panels of Figure \ref{fig_VB_img}, on the other hand, is much less pronounced. It is not shown here, but this lack of strong correlation persists even when the large-scale background $v$ and $B$ fields are removed.

\begin{figure}
\begin{center}
\includegraphics[width=7.5cm]{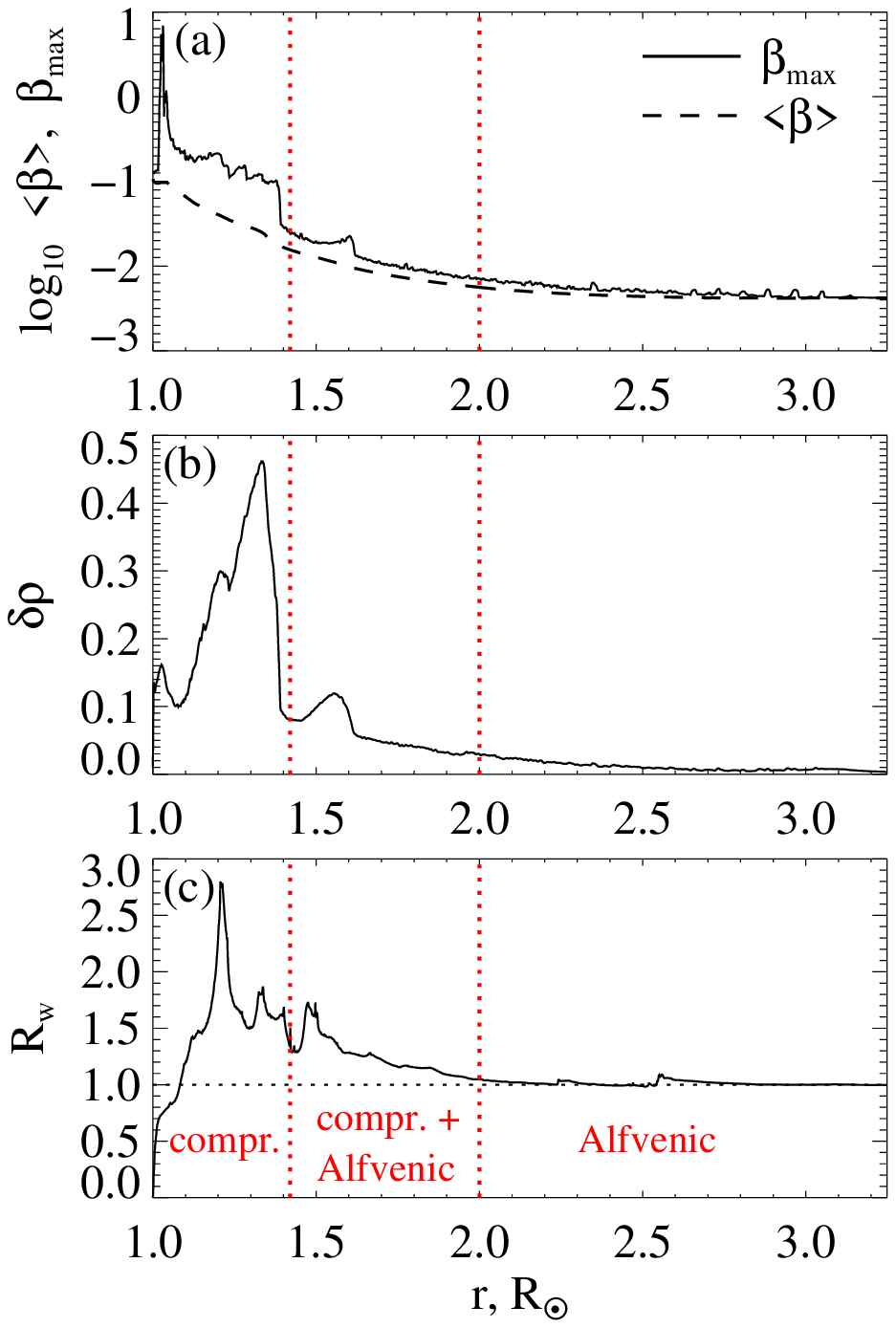}
\caption{\label{fig_plasma} Radial dependence of plasma parameters at $t = 3650$ s, revealing three distinct jet regions shaped by compressional and Alfv\'{e}nic perturbations and their combination: (a) average and maximum values of plasma $\beta$, (b) normalized mass-density fluctuation amplitude $\delta_\rho$ from Equation (\ref{eq:deltarho}), and (c) Wal\'{e}n ratio $R_w$ from Equation (\ref{eq:walen}).}
\end{center}
\end{figure}

\begin{figure}
\begin{center}
\includegraphics[width=7.4 cm]{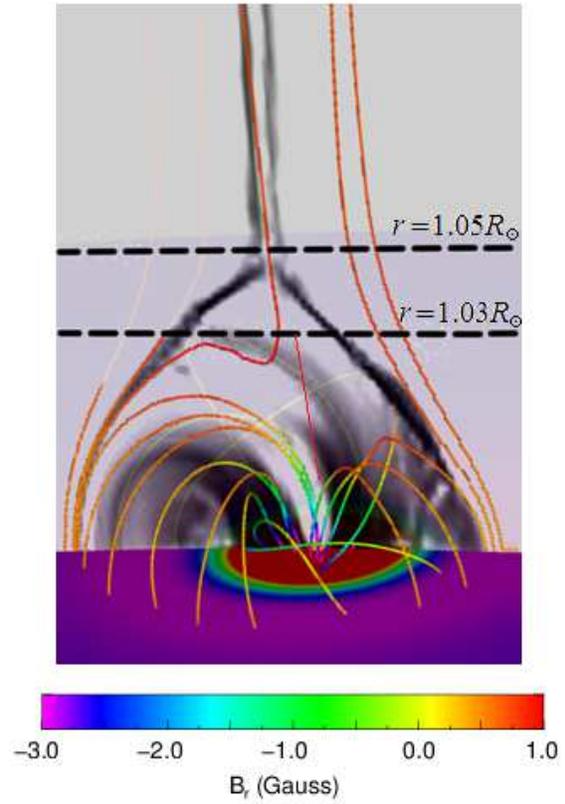} 
\caption{\label{fig_jet} Magnetic field lines in the low corona at $t = 3650$ s.  Color shadings indicate sign and strength of $B_r$ on the bottom surface and current-density magnitude $\vert J \vert$ in the plane of the sky.}
\end{center}
\end{figure}

Figure \ref{fig_plasma}a shows that the average value of plasma $\beta$ is small at all altitudes, signaling that magnetic pressure prevails over thermal pressure. Its maximum value $\beta_{\rm max}$, however, is significantly larger than the average and has substantial fluctuations at low altitudes $r < 1.4 R_\odot$. Beyond $r \approx 2 R_\odot$, on the other hand, the maximum $\beta$ value converges rapidly to the average value, indicating that the local perturbations become negligible, and both values are very small. The large spike in $\beta_{\rm max}$ near the photospheric boundary ($r \approx 1.03 R_\odot$) coincides with the location of the null region near the separatrix, as is evident from Figure \ref{fig_jet}. 

Figure \ref{fig_plasma}b shows the normalized mass-density fluctuation amplitude $\delta \rho$, which we define as the ratio of the standard deviation of the mass density $\sigma_{\rho}$ to its mean value $\left\langle \rho \right\rangle$,
\begin{equation}
\delta \rho \left( r \right) \equiv \frac{\sigma_{\rho}}{\left\langle \rho \right\rangle_{\theta,\phi}} = \frac{\left\langle \left( \rho - \left\langle \rho \right\rangle_{\theta,\phi} \right)^2 \right\rangle^{1/2}}{\left\langle \rho \right\rangle_{\theta,\phi}}.
\label{eq:deltarho}
\end{equation}
The averaging denoted by $\left\langle ... \right\rangle_{\theta,\phi}$ and the calculation of $\sigma_\rho$ are done over latitude and longitude at each radial distance $r$.  At $r < 2 R_\odot$, the mass density has a relatively large fluctuation amplitude which becomes particularly strong at At $r < 1.4 R_\odot$, showing that the mass density is quite variable in this inner region and that the flows are compressional. At $r > 2 R_\odot$, in contrast, the fluctuation amplitude approaches zero, indicating that in this outer region the flows are incompressible.

The \citet{walen1944} number $R_w$ is defined as the ratio of the transverse plasma flow speed to the Alfv\'en speed associated with the transverse magnetic field, 
\begin{equation}
R_{w} \left( r \right) \equiv \left\langle \frac{V_{\theta\phi}}{ B_{\theta\phi} / \left( \mu_0 \rho \right)^{1/2}} \right\rangle_{\theta,\phi}.
\label{eq:walen}
\end{equation}
For an ideal, linear or nonlinear Alfv\'{e}n wave in an isotropic plasma, $R_w = \pm1$.  As Figure \ref{fig_plasma}c shows, in our simulation $R_w > 1$ over the lower compressional region below $1.4 R_\odot$, where the plasma ejected by the reconnection events generates substantial nonuniformities in the mass density. Beyond $r \approx 2 R_\odot$, on the other hand, $R_w$ converges to 1 and remains there at higher altitudes. This indicates that the upper region is strongly dominated by Alfv\'{e}n waves. The intermediate region between $1.4$ and $2.0 R_\odot$ in which the Wal\'{e}n number is greater than one but density fluctuations are relatively small, is likely controlled by both compressible and shear Alfv\'{e}nic modes.

Taking the plots shown in Figure \ref{fig_plasma} together, it becomes clear that the jet combines three distinct regimes -- the compressional, the Alfv\'{e}nic, and  the mixture of the two. The dynamics occur close to the photosphere, and therefore the plasma structures are generated in the lower compressional region and then continue to propagate outwards with the jet velocity while maintaining their relative structure \citep{roberts15}. Consequently, even with gravity and solar wind taken into account, signatures of the low-altitude dynamics should be detectable by analyzing the jet structures present at the higher spacecraft altitudes.

\begin{figure*}
\begin{center}
\includegraphics[width=14cm]{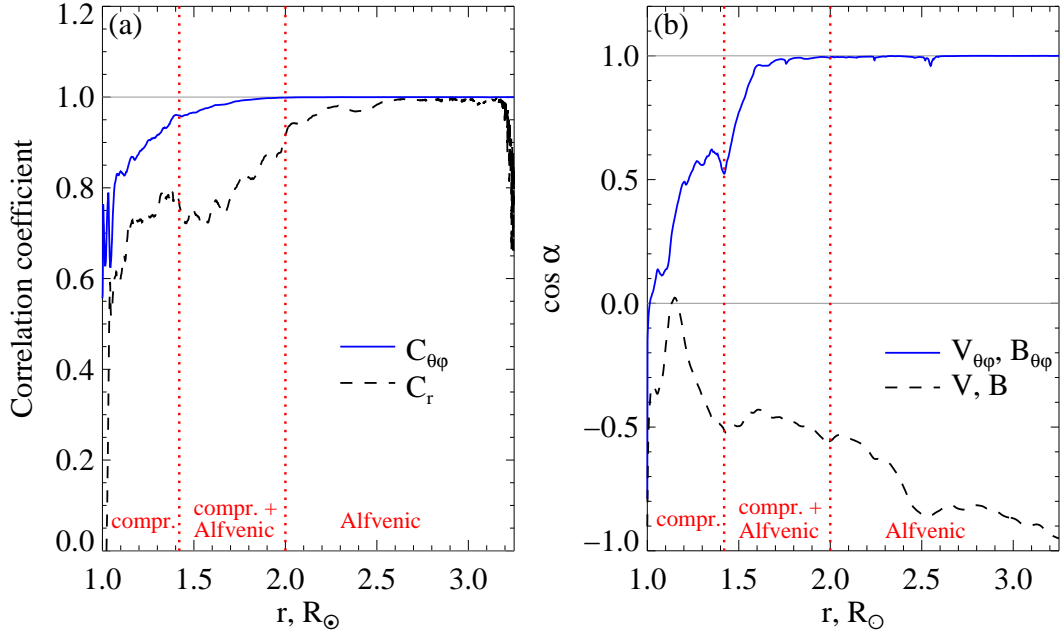}
\caption{\label{fig_corr} (a) Radial dependence of linear correlation coefficients between velocity and magnetic-field fluctuations in the transverse ($C_{\theta \phi}$) and radial ($C_r$) directions. (b) Radial dependence of average direction cosines between the two fields in the $\theta$-$\phi$ plane and in the 3D volume.}
\end{center}
\end{figure*}

The Alfv\'{e}nic nature of the velocity- and magnetic-field fluctuation amplitudes, whose transverse components are defined by Equation (\ref{deltavb}), can be tested conveniently by comparing their magnitudes and directions. Figure \ref{fig_corr}a presents the linear cross-correlation (Pearson) coefficients between their transverse components (coefficient $C_{\theta\phi}$) and their radial components (coefficient $C_r$). Both the transverse and radial coefficients are small near the lower boundary where $C_r$ drops to zero and $C_{\theta\phi}$ takes moderate values between 0.6 and 0.8, suggesting a partial decoupling of the velocity and the magnetic field in the active reconnection region. The correlations become stronger with height. At $r > 2 R_\odot$, $C_{\theta\phi}$ approaches 1 asymptotically, and the magnitudes of the transverse $v_{\theta\phi}$ and $B_{\theta\phi}$ perturbations become fully correlated as expected for a pure Alfv\'{e}n mode. The radial correlation coefficient drops above $3.2 R_\odot$ since the jet has not reached beyond this altitude by the studied time step. The direction cosine between the velocity and magnetic field perturbations in the $\theta,\phi$ plane (Figure \ref{fig_corr}b) is close to unity at and above the radial position $r \approx 1.8 R_\odot$. Below this altitude, there is significant misalignment between the transverse fields, suggesting the presence of compressional modes and more complex flow topologies. 

Figure \ref{fig_corr}b also shows that the total field vectors tend to become anti-parallel at $r > 2.5 R_\odot$. This reflects a well-ordered solar-wind outflow proceeding against the inward-directed magnetic field and constrained by cross-field magnetic pressure and tension forces prevailing in this region.


\section{Statistical Properties of Fluctuations}
\label{sec:stat} 

\subsection{Structure Functions}
\label{sec:SF}

\begin{figure*}
\begin{center}
\includegraphics[width=14cm]{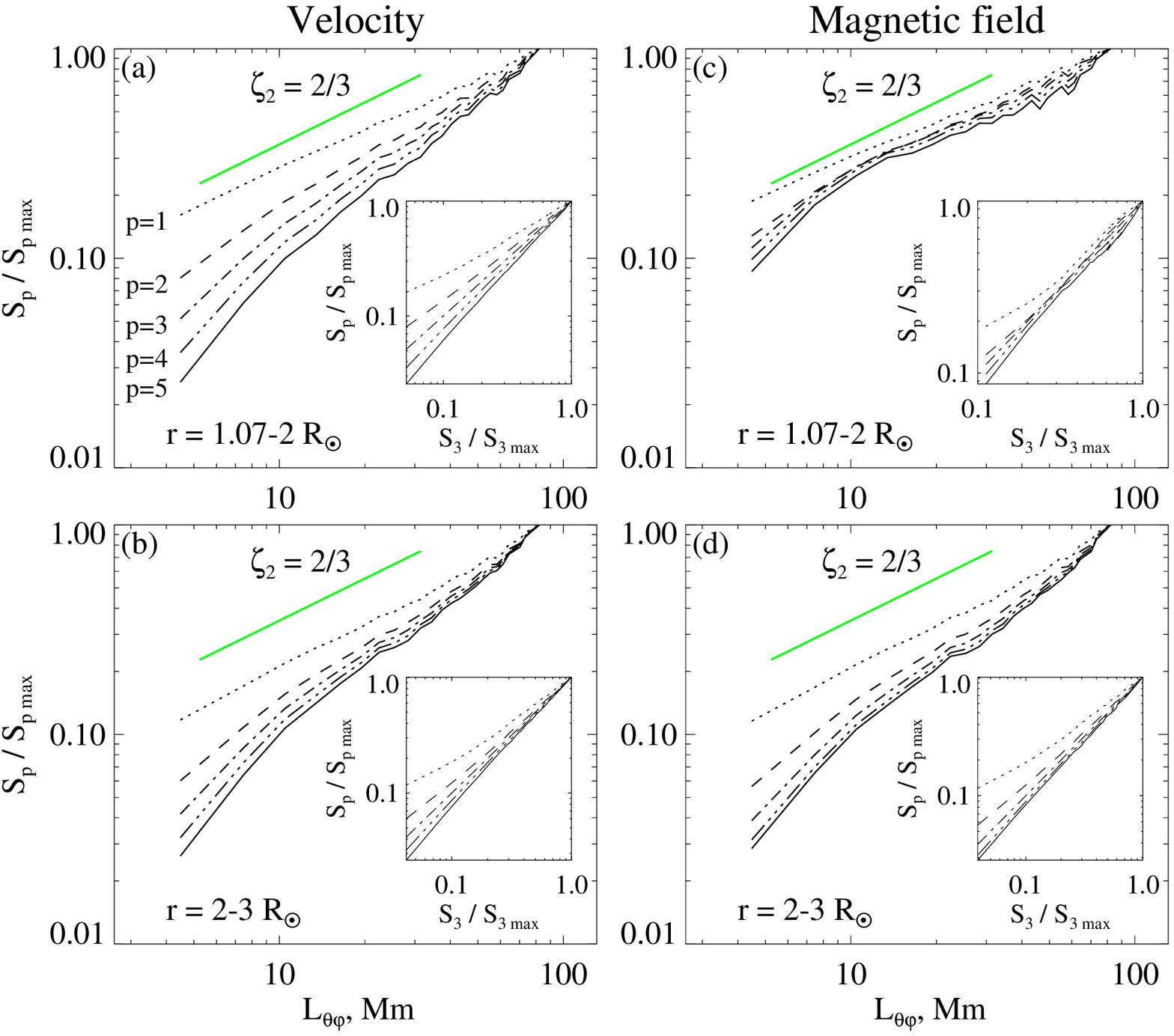}
\caption{\label{fig_SF} Structure functions of the transverse velocity (a,b) and magnetic (c,d) field fluctuations at $t = 3650$ s, below (top) and above (bottom) the altitude $r = 2R_\odot$ separating the compressional and Alfv\'{e}nic regimes, respectively. Main panels show the original SFs before applying the ESS normalization; solid green lines mark K41 slopes for comparison. Inset panels show the ESS-normalized SFs with improved inertial ranges.}
\end{center}
\end{figure*}

To determine the spatial scaling properties of the velocity and magnetic field perturbations,  we computed the unsigned scalar structure functions (SFs) 
\begin{equation}
S_p(r,L_{\theta\phi}) = \left\langle \left| f_{\theta\phi}(\bm{x_i},t)-f_{\theta\phi}(\bm{x_j},t) \right|^p  \right\rangle_{i, j},
\label{eq_SF}
\end{equation}
in which $f_{\theta\phi}$ is the magnitude of the transverse component of either field, $v_{\theta\phi}$ or $B_{\theta\phi}$, and the averaging denoted by $\left\langle ...\right\rangle$ is performed over all pairs of spatial positions $\bm{x_i}$ and $\bm{x_j}$ separated by the distance $L_{\theta\phi}$ (within a specified numerical accuracy) in planes of constant $r$, $(\bm{x_i}-\bm{x_j})\cdot\hat{r}=0$. We used integer-valued orders $p$ ranging from 1 to 5. 

For fully developed and adequately resolved turbulent flows, the empirical SFs in Equation (\ref{eq_SF}) assume a power-law form within the inertial range of scales. The log-log slopes of the SFs estimated in this range serve as empirical proxies for the theoretical exponents $\zeta_p$ describing the higher-order scaling of transverse $v_{\theta\phi}$ and $B_{\theta\phi}$ perturbations, as discussed in \S \ref{sec:turb}. The slope of the second-order SF has special significance as it yields the exponent $\zeta_2$ directly related to the slope $\alpha=\zeta_2+1$ of the Fourier power spectrum of the fluctuations \citep[see, e.g.,][]{uritsky11}.

If the range of scales of the intrinsic turbulent dynamics is not adequately resolved due to limited data resolution, the SF scaling may exhibit significant non-power law distortions. To overcome this problem, the method of extended self-similarity \citep[ESS;][]{benzi93} is often used. ESS allows the observable range of turbulent scaling to be extended by exploiting the dependence of $S_p$ on the third-order SF, which is expected to obey the relation $S_p \approx S_3^{\zeta_p/\zeta_3 }$. This power-law scaling of $S_p$ vs.\ $S_3$ typically spans a broader dynamic range than the direct scaling of $S_p$ vs.\ $L_{\theta\phi}$. Consequently, a more accurate estimation of the relative exponents $\xi_p=\zeta_p/\zeta_3$ results \citep{muller00,uritsky07}, even when the observed inertial range in $L_{\theta\phi}$ is small. In hydrodynamic turbulence models, $\zeta_3=1$, a constraint that follows from the Navier-Stokes equations \citep[see, e.g.,][]{landau87}. Therefore an ideal non-magnetized fluid should show $\zeta_p = \xi_p$ $\forall$ $p$, justifying the third-order normalization $\xi_p=\zeta_p/\zeta_3$.

Figure \ref{fig_SF} presents a set of velocity and magnetic-field SFs averaged over two ranges of radial coordinates, above and below $2 R_\odot$, at time $t=3650$ s. By this time, the jet front has reached altitude $r \approx 3.5 R_\odot$ and a pronounced turbulent wake trails below it. For the range of radial coordinates shown in the figure, the grid spacing of the interpolated data arrays was 0.001 $R_\odot$, 0.057$^{\circ}$ and 0.028$^{\circ}$ in the $r$, $\theta$, and $\phi$ directions, respectively. We excluded the region $r / R_\odot > 3$ from the analysis due to its lower resolution, which was inadequate for the SF calculations. Tilted solid lines (green) show the K41 values of the $\zeta_2$ exponent in each panel of Figure \ref{fig_SF}, for reference. 

The velocity SFs computed for both ranges of $r$ (Figures \ref{fig_SF}a,b) exhibit the systematic increase of the log-log slope with order $p$ typical of turbulent fluids. The slopes saturate for $p > 3$, suggesting that non-uniform dissipation introduces KRS corrections to the linear $\zeta_p$ dependence in Equation (\ref{eq_KRS}). This tendency also can be seen in the SF plots of magnetic-field fluctuations above $r = 2 R_\odot$ (Figure \ref{fig_SF}d), but not at lower altitudes (Figure \ref{fig_SF}c), where the SFs associated with different $p$ nearly collapse onto a single curve. This behavior signals the presence of extremely intermittent structures that dominate the statistical averaging \citep{she94}. It is likely that these singular structures are associated with intense current sheets formed by the reconnection of field lines at the dome and stretched out by the upward plasma flow. 

The insets in Figure \ref{fig_SF} show the ESS-transformed structure functions. Most of these are considerably closer to straight lines on the double-logarithmic scale than the original SFs provided in the main plots. This suggests that some of the non-power-law distortions seen in the original SFs are caused by insufficient grid resolution, and the ESS exponents $\xi_p$ should be a more reliable marker of the underlying multiscale structures.

Overall, the shapes of the averaged velocity and magnetic field SFs shown in Figure \ref{fig_SF} are rather similar for $r > 2 R_\odot$. This suggests that the transverse $v_{\theta\phi}$ and $B_{\theta\phi}$ perturbations are strongly coupled through the mean magnetic field across the entire range of spatial scales analyzed. This coupling is essentially lacking at lower jet altitudes, however, where compressional modes and the reconnection process affect the velocity and magnetic field SFs differently.

\subsection{Radial Dependence of SF Exponents}
\label{sec:zetas} 

To investigate the jet turbulence at $t=3650$ s further, we evaluated the $\zeta_p$ exponents within an inertial range of scales $L_{\theta\phi}$ between $\sim$5 and $\sim$30 Mm. The lower boundary is the resolution limit; it corresponds to the average angular separation of about 0.2$^{\circ}$, which is twice the transverse spacing of the regularized grid averaged over $r \in [1 R_\odot,3 R_\odot]$. The upper limit of the studied range reflects the characteristic angular scale ($\sim 1.0^{\circ}$) of the largest $v_{\theta\phi}$ and $B_{\theta\phi}$ perturbations embedded into the jet (see Figure \ref{fig_VB_img}).

For each set of SFs, the chosen inertial range was also used to determine the range of $S_p/S_3$ values used to compute the ESS exponents $\xi_p$.

\begin{figure*}
\begin{center}
\includegraphics[width=14cm]{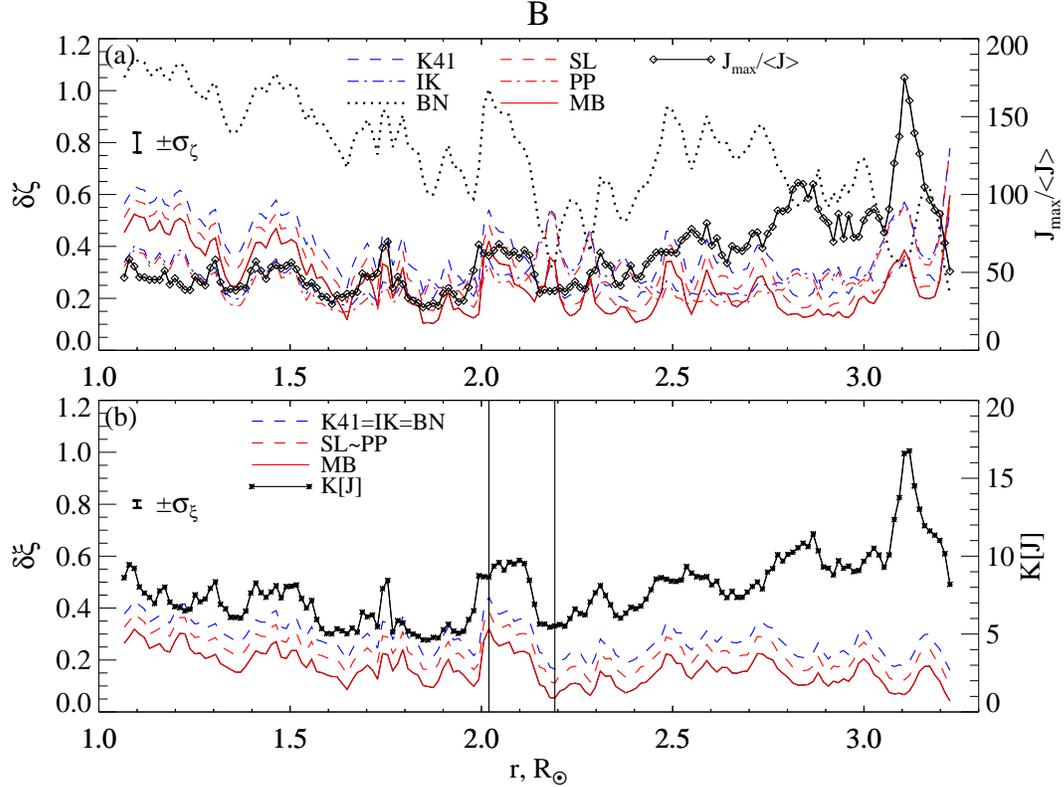}
\caption{\label{fig_rad} Radial dependence of root-mean-square discrepancies between measured and theoretical values of the magnetic-field SF exponents, $\delta \zeta$ and $\delta \xi$, defined by Equation (\ref{eq:deltazetaxi}). Exponents are calculated using the (a) original and (b) ESS-transformed structure functions for several hierarchical models of intermittent turbulence described by Equation (\ref{eq_hier}) and discussed in \S \ref{sec:turb}. By definition, zero rms discrepancies mean a perfect match with a particular model. Vertical lines mark radial positions with the lowest and highest discrepancies for $r < 3R_\odot$. $J_{max}/ \left\langle J \right\rangle$, the ratio of the maximum to the average total current density, and the current density kurtosis $K[J]$, are added as indicators of current intermittency at each altitude. Average standard errors of $\zeta$ and $\xi$ estimates (respectively $\sigma_\zeta$ and $\sigma_\xi$) are shown with vertical straight line segments in the left-hand side portion of each panel.}
\end{center}
\end{figure*}

Figure \ref{fig_rad} presents the results of the comparison of magnetic-field SF exponents describing the jet with several statistical models of turbulence described in \S \ref{sec:turb}: the non-intermittent fluid and MHD cascade models (K41 and IK), their intermittent counterparts (SL and PP), the MB model combining a hydrodynamic-like cascade with intermittent dissipative MHD structures, and the simple BN model with its fully uncorrelated spatial increments.

The performance of each model has been evaluated using the $p$-averaged root-mean-square (rms) discrepancies $\delta \zeta$ and $\delta \xi$,
\begin{align}
\delta \zeta &\equiv \left\langle \left( \zeta_p - \zeta_p^{th} \right)^2 \right\rangle_{p} ^{1/2},\nonumber \\
\delta \xi &\equiv \left\langle  \left( \xi_p - \xi_p^{th} \right)^2 \right\rangle_{p} ^{1/2}.
\label{eq:deltazetaxi}
\end{align}
Here, $\zeta_p^{th}$ and $\xi_p^{th}$ are the theoretical exponents and $\left\langle...\right\rangle_{p}$ denotes averaging over $p = 1,...,5$. Smaller discrepancies imply greater accuracies in the associated theoretical model. 

Figure \ref{fig_rad}a shows the radial dependence of the rms discrepancies of the non-ESS transformed $\zeta_p$ exponents. The discrepancy values produced by the BN model are by far the highest in the group. This indicates that the assumption of uncorrelated fluctuations underlying the simplistic BN model is inconsistent with the dynamics of the jet. For $r < 2 R_\odot$, the hydrodynamic models K41 and SL based on viscous dissipation tend to show the next-largest discrepancies after the BN model. The IK and PP models track each other closely; they and MB show mixed performance without an obvious winner. For $r > 2 R_\odot$, the discrepancy of the MB model becomes systematically lower than that of any other model.

The rms discrepancies of the ESS-transformed $\xi_p$ exponents shown in Figure \ref{fig_rad}b confirm this tendency, and suggest that the MB fit is the best both below and above $r = 2 R_\odot$. Note that since all non-intermittent models (i.e.\ K41, IK, and BN) predict linear growth of $\zeta_p$ with $p$, their ESS exponents $\xi_p \equiv \zeta_p/\zeta_3$ are indistinguishable from one another. Intermittent models, however, are clearly distinguishable from one another and from the non-intermittent models based on ESS as evident from the Figure.

All rms indicators exhibit considerable variability across the range of distance from the solar surface. For the lower portion of the jet, this variability positively correlates with electric-current intermittency as expressed by $J_{max}/\left\langle J \right\rangle_{\theta,\phi}$, the ratio of the largest to the mean value of the current density magnitude at a given $r$ (solid black curve with diamonds on Figure \ref{fig_rad}a), as well as the excess kurtosis of the current
\begin{align}
K[J] = \frac{ \left\langle \left( J - \left\langle J \right\rangle_{\theta,\phi} \right)^4 \right\rangle_{\theta,\phi}}{\left\langle \left(J-\left\langle J\right\rangle_{\theta,\phi} \right) ^2 \right\rangle_{\theta,\phi}^2 } - 3
\label{eq:kurtosis}
\end{align}
shown with asterisks on Figure \ref{fig_rad}b. The Pearson correlation coefficient between the current kurtosis 
and the $\zeta_p$ ($\xi_p$) discrepancy for the MB model reaches 0.60 (0.73) in the region $r \in [1.3 R_\odot, 2.3 R_\odot]$.  Statistically, the large positive excess kurtosis and the increased max-to-mean ratio of current density fluctuations are associated with a development of a heavy distribution tail making extreme current density values more probable \citep{kinney95}, compared to the normal distribution for which $K=0$.

\begin{figure}
\includegraphics[width=7.5cm]{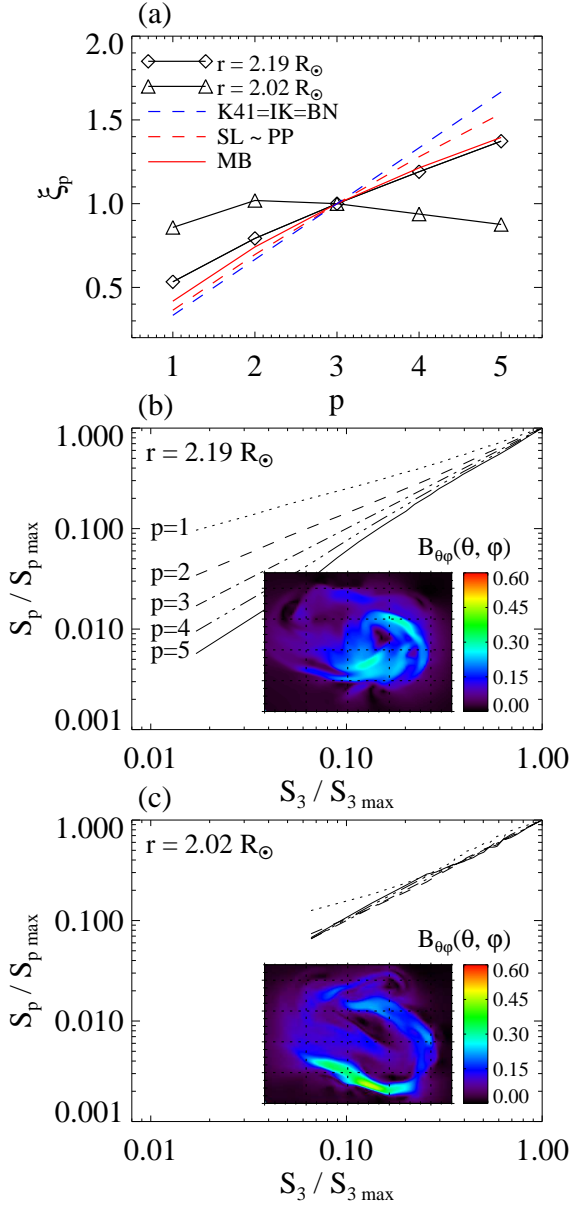}
\caption{\label{fig_SF_cuts} (a) ESS SF exponents of transverse magnetic field fluctuations for two characteristic radial positions (shown in Figure \ref{fig_rad}), demonstrating the lowest and highest discrepancies with the theoretical models. (b,c) ESS-transformed SF functions and maps of transverse magnetic-field magnitudes for the two positions.}
\end{figure}

We found that the non-Gaussian current enhancements leading to less accurate theoretical SF exponent predictions are caused by intense current sheets. This explanation is demonstrated in Figure \ref{fig_SF_cuts}, which compares magnetic field turbulence at two radial positions ($2.02 R_\odot$ and $2.19 R_\odot$) marked by vertical lines in Figure \ref{fig_rad}b. They are the positions of, respectively, the highest ($\approx 0.32$) and lowest ($\approx 0.05$) MB discrepancies for the magnetic field fluctuations. 

The dependence of the ESS SF exponents on the order $p$ is drastically different at these two locations (Figure \ref{fig_SF_cuts}a). At $r = 2.19 R_\odot$, the dependence is reasonably close to the hierarchical scaling predicted by the MB model. This agreement, as well as the distinctions with the non-intermittent models, is especially clear at higher $p$. Since higher SF orders are more sensitive to intermittency effects, the convergence is expected provided that MB phenomenology captures the jet physics. At $r = 2.02 R_\odot$, $\xi_p$ reaches its maximum at $p=3$, after which it begins to decrease. This contrasts strikingly with all theoretical models, both homogeneous and intermittent, which show monotonic growth with $p$. The inverted hierarchy of SF exponents ($\partial \xi/\partial p <0$) is suggestive of a singular high-amplitude disturbance embedded in a stochastic background \citep{uritsky11}. 

The spatial patterns of $B_{\theta\phi}$ perturbation shown in the Figures \ref{fig_SF_cuts}b,c insets visualize the difference between the two jet conditions. The magnetic field fluctuations at $r = 2.19 R_\odot$ exhibit multiple spatial scales indicative of a turbulent flow, in agreement with the hierarchy of SFs presented on the main panel. In contrast, magnetic field variability at $r = 2.02 R_\odot$ is much more ordered compared to the surrounding flow. Magnetic fluctuations at this altitude are dominated by a single monoscale high-intensity magnetic structure leading to a collapse of the higher-order SFs onto the same curve, which is inconsistent with the hierarchical turbulence models.

\subsection{Temporal Dependence of SF Exponents}
\label{sec:temp} 

\begin{figure*}
\begin{center}
\includegraphics[width=12 cm]{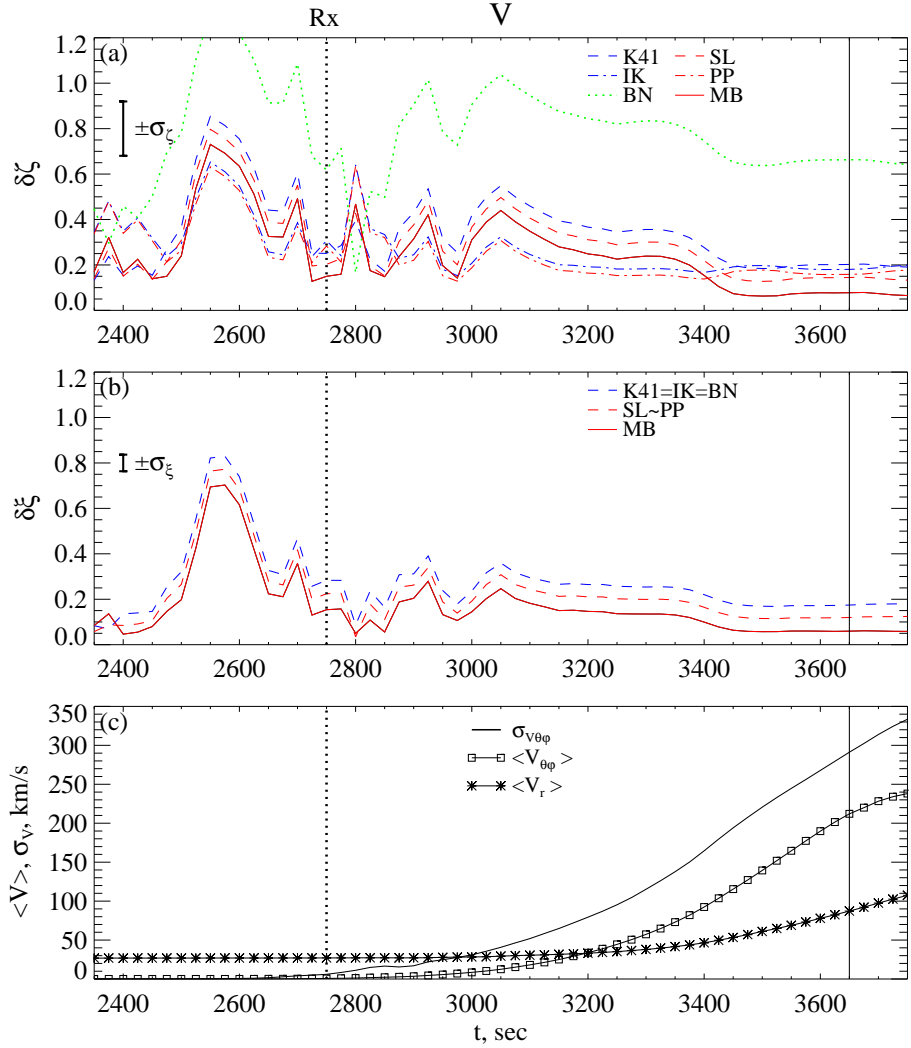}
\caption{\label{fig_temp}Temporal evolution of root-mean-square discrepancies between measured and theoretical values of velocity SF exponents. Exponents are calculated using the (a) direct and (b) ESS-transformed structure functions, for several hierarchical models of intermittent turbulence defined by Equation (\ref{eq_hier}) and discussed in \S \ref{sec:turb}. For comparison, the average radial and transverse velocity magnitudes and the standard deviation of the transverse velocity at $\phi=0$ are displayed in panel (c). Dotted vertical lines indicate reconnection jet onset. The vertical lines at $t = 3650$ s mark the time discussed in Section \ref{sec:zetas}.}
\end{center}
\end{figure*}

Figure \ref{fig_temp} compares the time evolution of the transverse velocity fluctuation amplitudes $v_{\theta\phi}$ within the central two-dimensional radial slice ($\phi=0$) with the commonly used theoretical models discussed in \S \ref{sec:turb}. Figure \ref{fig_temp}a shows the root-mean-square (rms) discrepancy $\delta \zeta$ defined by Equation (\ref{eq:deltazetaxi}) between the measured ($\zeta_p$) and theoretical ($\zeta_p^{th}$) SF exponents. Figure \ref{fig_temp}b presents similar discrepancy plots for the ESS-normalized exponents $\xi_p = \zeta_p/\zeta_3$. Average standard errors of $\zeta$ and $\xi$ estimates (respectively $\sigma_\zeta$ and $\sigma_\xi$) are shown with error bars on both panels for comparison. Figure \ref{fig_temp}c shows the time evolution of mean values of the radial and transverse velocity components, as well as the standard deviation of the transverse  velocity. 

The discrepancies of the original (non ESS-normalized) exponents shown in Figure \ref{fig_temp}a are marked by significant variability during the $\approx 500$ s interval centered on the reconnection onset at $t \approx 2750$ s (vertical dashed lines), revealing non-stationary plasma conditions during this time. The variation of the exponents before the reconnection onset is not physically meaningful because it was accompanied by near-zero  mean value and standard deviation of the transverse velocity (see Figure \ref{fig_temp}c). Turbulence models are not appropriate to this early phase of the evolution. After the reconnection onset, there is a transition from rapid fluctuations in the exponents to a quasi-steady phase as the turbulence develops and becomes ever stronger.

Starting from $t = 3050$ s, the discrepancies in all of the models generally decrease. It is important to note that the three hydrodynamic cascade models (K41, SL, and MB) show a more pronounced discrepancy decrease compared to the MHD cascade models (IK and PP). After $t \sim 3400$, when the turbulence in the jet wake becomes fully developed, the MB model demonstrates the best performance, characterized by the lowest discrepancy ($\approx 0.07$) among the studied group of models by the end of the run. 

The Brownian noise model is inconsistent with the jet behavior, except during a short transient drop soon after the reconnection onset. The inadequacy of the BN model indicates that the velocity fluctuations are rooted in the nonlinear dynamics of the physical fluid rather than, for example, possible numerical effects that produce uncorrelated errors. 

The normalized discrepancy plots presented in Figure \ref{fig_temp}b suggest that, in fact, the MB model outperformed all other models during the entire jet evolution. In the ESS representation, the three non-intermittent models (K41, IK, and BN) are indistinguishable. They performed equally poorly compared to the other models, signaling that the turbulent velocity field contains coherent intermittent structures that significantly modify its cascade dynamics. The intermittent hydrodynamic models (SL and PP) performed better, but not as well as the intermittent MHD model (MB). For all models, the decrease of the rms discrepancies is accompanied by a steady growth of the mean radial velocity, as shown in Figure \ref{fig_temp}c. This joint evolution can be interpreted as a transition from a partially- to fully-developed turbulent state driven by the plasma flow as it propagates to higher altitudes, with the turbulent wake region downstream of the main jet front providing the best conditions for the intermittent cascade.

The vertical solid lines added to Figure \ref{fig_temp} mark the time $t = 3650$ s examined in the previous section. Around this time, the MB discrepancy stabilized at its lowest steady level throughout the run. This indicates that the MB-type cascade was well established, in agreement with our spatial analysis presented above.





\section{Multiscale Current Structures}
\label{sec:cs}

\begin{figure*}
\begin{center} 
\includegraphics[width=12cm]{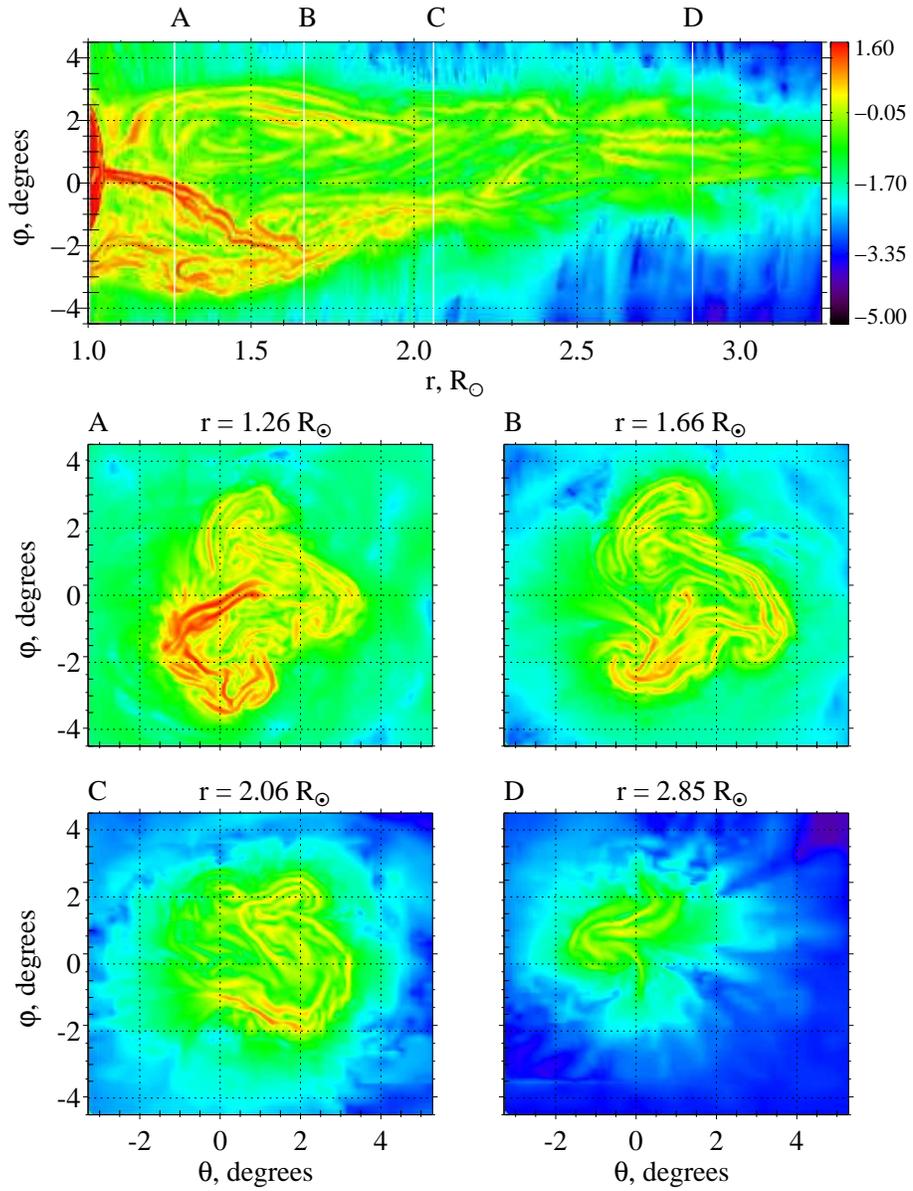}
\caption{\label{fig_J_cuts} Constant-latitude and constant-radius cross sections of the logarithmic current density magnitude, revealing multiple current sheets of various sizes and complex shapes. White lines labeled A-D in the top image indicate the radial locations of the cross-sections in the lower images.}
\end{center} 
\end{figure*}

The intense magnetic structures perturbing the multiscale hierarchy of SFs (Figure \ref{fig_SF_cuts}) are associated with strong currents. This is evident from Figure \ref{fig_J_cuts}, which shows longitudinal and transverse cross sections of the $\text{log}_{10} J $. We used the logarithmic intensity scale to expand the dynamic range of the color table, which allows us to visualize both the strongest singular current structures and the significantly weaker multiscale current fluctuations generated by the jet turbulence. The plasma regions carrying the strongest currents below $r \approx 1.6 R_\odot$ are magnetically connected to the separatrix surface, which is not plotted here, and consist of reconnected magnetic field lines. The surrounding structures are quite chaotic and exhibit scale and curvature variability at all scales. The current morphology becomes simpler and more uniform with altitude, but the current continues to be spread over a wide range of spatial scales.

A visual comparison of the radial and transverse cross sections in Figure \ref{fig_J_cuts} suggests that the current structures take the form of radially elongated current sheets that are arranged in a distorted quasi-cylindrical pattern imposed by the bulk plasma flow at the largest scales. We tested the geometry of such embedded turbulent current structures quantitatively using a cluster detection method as follows.  


\begin{figure*}
\begin{center} 
\includegraphics[width=15cm]{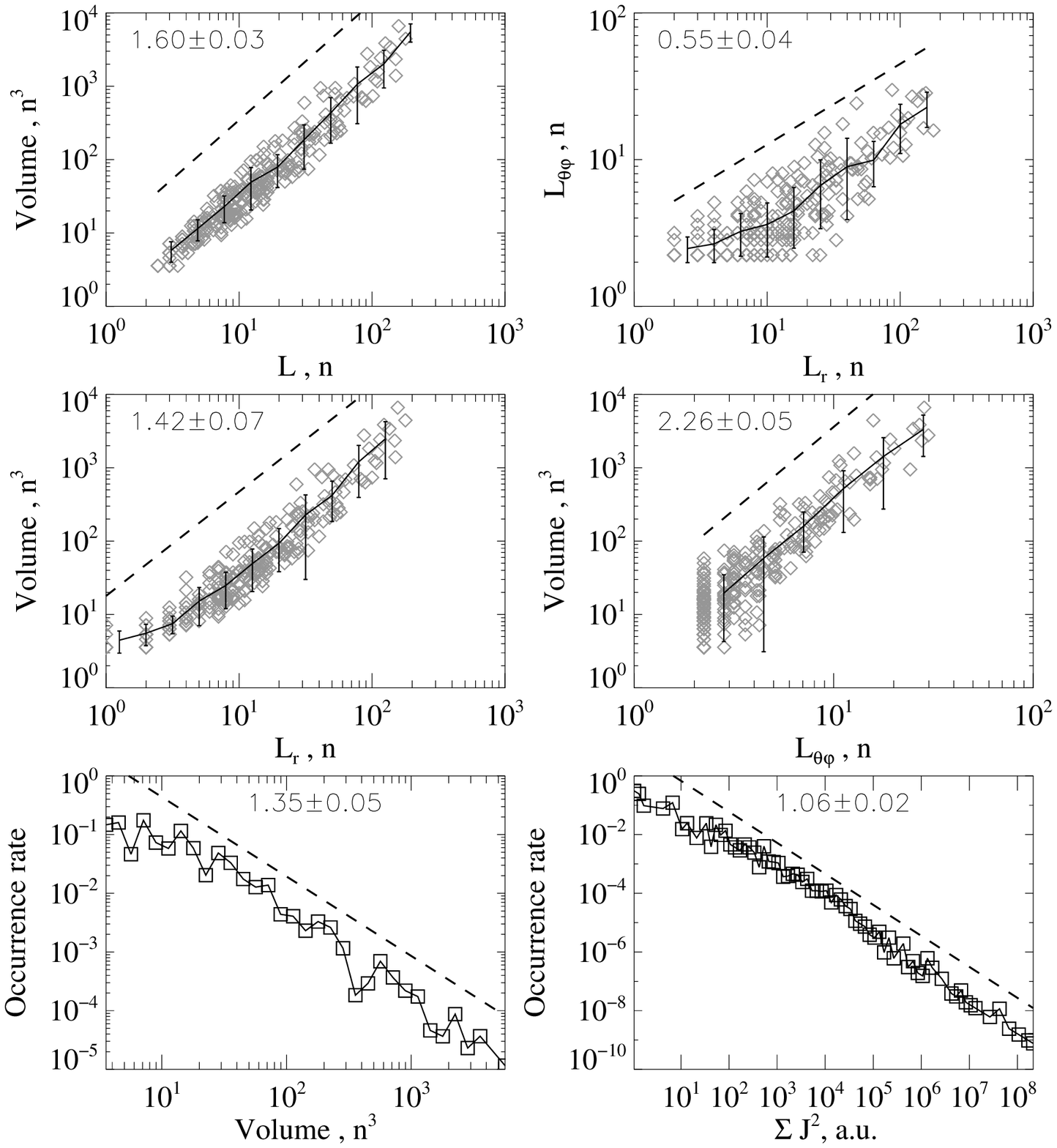} 
\vspace{10mm}
\caption{\label{fig_CS_2} Scatter plots and distribution functions of several parameters of turbulent current sheets inside the simulated jet revealing their multiscale anisotropic structure
(see text for details).}
\end{center} 
\end{figure*}

Turbulent flows exhibit small-scale structures, with strong gradients, where dissipation takes place. In principle, these structures can be detected through any relevant physical variable, but the most relevant to turbulent MHD flow is the current. Our analysis therefore focused on the 3D array containing the values of $J$ at $t=3650$ s. A grid node is considered to belong to a small-scale structure if the amplitude in this node, expressed in terms of $J$, exceeds the level of $m$ standard deviations above the mean value for a given radial position $r$:
\begin{equation}
\begin{split}
J_{th}(r) & \!=\! \left\langle J(\theta,\!\phi,\!r) \right\rangle_{\theta,\phi} \\
          & \!+\!  m \left( \left\langle J^2(\theta,\!\phi,\!r) \right\rangle_{\theta,\phi}\! -\! (\left\langle J(\theta,\!\phi,\! r) \right\rangle_{\theta,\phi})^2\right)^{1/2}.\\
\label{eq_thresh}
\end{split}
\end{equation}
Intermittent structures in the $J$ field are defined as spatially connected sets of grid nodes satisfying the threshold condition $J > J_{th}$. 
In this way, we identify contiguous spatial regions that are anticipated to exhibit enhanced Joule dissipation and examine their geometry across the inertial range of scales. We used the cluster detection algorithm described by \citet{uritsky10, uritsky10a} to separate these structures from the background and to ascertain their individual properties, such as linear size, volume, and inferred dissipation rate.

To overcome the memory limitations of standard cluster-detection algorithms, we applied an optimized technique that first identified the grid nodes belonging to the clusters and then traced their spatial connectivity. First we built the {\it activation table} \citep{uritsky10a}: a table of contiguous intervals along the radial direction, used here as the scanning direction, where the  current-density magnitude exceeds the detection threshold. Next we found and labeled spatially connected clusters of activations using the ``breadth-first search'' principle \citep{lee61} to avoid repeatedly reconstructing the search trees representing individual clusters. We found it necessary to consider all 26 nearest neighbors in a $3 \times 3 \times 3$ array centered on each grid node, including along diagonals, when identifying connected activations. Finally, the activation table was sorted according to the cluster labels to enable fast access to the detected structures. The output data array preserves complete information on the location and shape of all the contiguous regions in the simulation volume where the threshold condition is fulfilled.


To obtain an ensemble-based statistical portrait of the detected current clusters, each was characterized by its volume $V$ (the total number of grid nodes involved in the cluster), the linear scales $L_r$ and $L_{\theta\phi}$ describing the spatial extent of the cluster along the radial and transverse directions respectively, the isotropic scale $L$ defined by 
\begin{equation}
L \equiv \left( L_{\theta\phi}^2 + L_r^2 \right)^{1/2},
\end{equation}
and the squared total current density $\vert J^2 \vert$, used as a proxy for the volume-integrated Joule dissipation rate.

The statistics obtained, shown in Figure \ref{fig_CS_2}, confirm the quasi-two-dimensional geometry of the turbulent current clusters. The scaling of volume versus the isotropic scale $L$ in Figure \ref{fig_CS_2}a exhibits a power-law dependence $V \propto L^{D_V}$ across a wide range of cluster sizes. The best-fit value of the $D_V$ exponent rules out the possibility that current clusters are geometric objects of dimension $D_V = 3$. We find that the actual morphology of the cluster is represented by a continuous distribution from long, quasi-1D filaments to small, quasi-2D sheets, culminating in $D_V < 2$. The statistical correlation between the transverse and radial cluster scales in Figure \ref{fig_CS_2}b shows that $L_{\theta\phi} \propto L_r^{0.67}$ and therefore the aspect ratio $L_{\theta\phi}/L_r$ scales as $L_r^{-0.33}$. This implies that the aspect ratio of the clusters decreases gradually with the radial length of the cluster, making the largest current sheets effectively one-dimensional but preserving the quasi-two-dimensional geometry of smaller current clusters, although all of them are to some degree anisotropic. 

The volume scaling is substantially different in the directions parallel and perpendicular to the bulk outflow as shown in Figures \ref{fig_CS_2}c and \ref{fig_CS_2}d. This anisotropy is described by the power laws $V \propto L_r^{1.42}$ and $V \propto L_{\theta\phi}^{2.26}$, yielding $L_{\theta\phi} \propto L_r^{0.63}$. This is very close to the $0.67$ scaling exponent of the directly measured correlation between $L_{\theta\phi}$ and $L_r$.


The bottom panels of Figure \ref{fig_CS_2} show the probability distributions of current-sheet volumes (Figure \ref{fig_CS_2}e) and volume-integrated $J^2$ measures (Figure \ref{fig_CS_2}f). Both histograms are power-laws with  indices less than 1.5, suggesting that the mean volume of the clusters and the net energy dissipation rate provided by clusters of a given size are controlled by the largest structures, even though small current sheets are much more abundant. 











\section{Discussion and Conclusions}
\label{sec:disc_conc} 

Our analysis of adaptively refined 3D MHD simulations confirms the occurrence of reconnection-driven turbulence in a supersonic coronal hole jet, and explores its structure and evolution. 

We have found that spatial correlations of magnetic fluctuations inside the jet are in quantitative agreement with a scaling ansatz of intermittent MHD turbulence proposed by \citet{muller00}. The MB scaling model implies that the turbulent cascade inside the jet is supported by filamentary structures representing fluid vortices, whereas the energy dissipation takes place in intermittent current sheets. The current sheets in the simulated coronal jet obey this scenario and exhibit a scale-dependent geometry, with the largest sheets stretched into highly elongated structures parallel to the large-scale flow and the smallest sheets being more isotropic. 

\begin{figure}
\begin{center} 
\includegraphics[width=7.9cm]{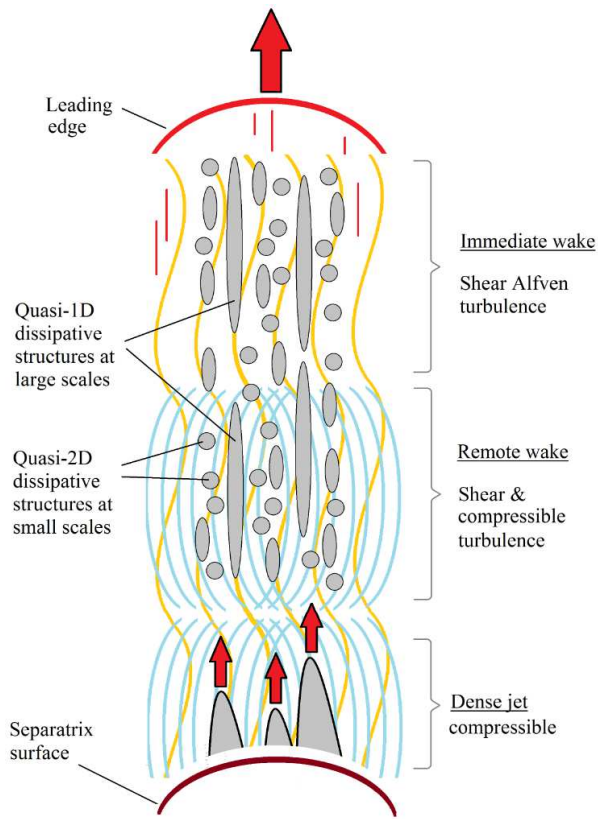} 
\vspace{10mm}
\caption{\label{fig_jet_sketch} Schematic diagram showing the internal structure of the jet according to our analysis. Regions of Alfv\'{e}nic and compressible wave activity are marked with yellow and blue curves, correspondingly.}
\end{center} 
\end{figure}

Our results show that the turbulent wake of the jet contains three radially stratified regions (Figure \ref{fig_jet_sketch}): (1) the immediate wake behind the leading edge of the jet, dominated by shear Alfv\'{e}n turbulence; (2) the remote wake characterized by both Alfv\'{e}nic and compressible turbulence; and (3) the dense portion of the jet adjacent to the reconnection driver and dominated by compressible, non-Alfv\'{e}nic plasma motions. Our additional analysis (not shown) indicates that the dense region is dynamically rather important as it carries the largest momentum and kinetic energy densities compared to the other two jet regions. 

The prevailing role of Alfv\'{e}nic turbulence in the immediate wake is confirmed by the Wal\'{e}n ratio $R_w \approx 1$ and by the spatial alignment of the transverse velocity and magnetic field perturbations, which maintain their multiscale structure as the jet propagates. The remote wake, in turn, exhibits a considerable degree of compression in addition to Alfv\'{e}nic fluctuations found in the immediate wake. The nature of stochastic motions in the dense jet remains to be understood. It is likely that the 3D geometry of the reconnected field plays a major role in this region but how exactly it couples to the multiscale plasma flow is not clear.

Coronal hole jets are most commonly observed at the Sun's polar regions. These jets can couple directly to the fast solar wind, where the lack of significant dynamical interaction between flows at different speeds creates conditions for an unperturbed propagation of extremely  low frequency Alfv\'{e}nic fluctuations across large heliocentric distances \citep[see, e.g.,][and references therein]{bruno13}. Most of what is currently known about turbulence in the polar wind is based on the observations by the Ulysses spacecraft in 1994-1995. These observations show that polar Alfv\'{e}nic turbulence evolves similarly to solar wind in the ecliptic plane but on a significantly slower time scale, due to the absence of strong velocity shears and interplanetary shocks \citep[e.g.,][]{bavassano00}. 

Ulysses magnetic field and plasma measurements above $\pm 30^\circ$ latitude demonstrate the presence of strong statistical correlations between the transverse components of magnetic and velocity fields characteristic of Alfv\'{e}nic turbulence \citep{smith95, goldstein95}. The spectral index of magnetic field variations in the polar wind is close to $\alpha = 5/3$ at frequencies above $\sim 10^{-3}$Hz \citep{horbury95, horbury96}. Within statistical uncertainty, these estimates are indistinguishable from the prediction of the MB model $\alpha \approx 1.741$ and the flows in the immediate wake of our jet. Higher order structure-function analysis of magnetic field fluctuations above the Sun's polar coronal holes \citep{nicol08} provides extra support for the MB scenario showing a stable exponent ratio $\zeta_2/\zeta_3 \approx 0.75$ consistent with Equation (\ref{eq_MB}).
 
For a nominal flow speed of the order of 700 km s$^{-1}$ and the lowest frequency $\sim 10^{-3}$Hz, the radial scales of the MHD turbulence in the polar wind should be no larger than $7 \times 10^5$km. Assuming that a volume of plasma transported by the polar wind expands in the transverse directions but not along the radial direction \citep{dong14}, this upper scale limit can be compared with jet fluctuations. Figure \ref{fig_VB_img} shows that the largest radial scale of jet fluctuations is on the order of $10^5$ km, which agrees with polar wind measurements. 

To summarize, observations suggest that the polar wind turbulence is dominated by the same type of energy cascade as the one found in the coronal hole jet simulation studied here. It should be noted that the compact size and the transient nature of coronal hole jets make it difficult to trace their individual contributions to the solar wind dynamics. However, a cumulative effect of many such events organized into relatively large and long-lived formations such as coronal plumes could strongly influence magnetic and velocity field fluctuations in the adjacent heliosphere and explain much of its stochastic structure. 

The upcoming {\it Solar Probe Plus} and {\it Solar Orbiter} solar missions will likely provide more insight into the physics of the jet turbulence and its relevance to the solar wind. Among indicative single-spacecraft turbulence tests that could be conducted by these missions are the higher-order time-domain structure functions converted into the spatial domain using appropriate dispersion relations, the Wal\'{e}n ratio and the velocity - magnetic field orientation analysis as local markers of shear-Alfv\'{e}n modes, and compressibility analysis using mass density fluctuations. In this context, additional simulations involving multiple coronal hole jets could be instrumental for clarifying the mechanism of the coupling of the jet flows with the fast wind. The fact that turbulent characteristics of a { \it single} jet agree with polar solar wind measurements, as has been established here, may indicate that a limited number of jets can in fact be sufficient to reproduce a realistic turbulent solar wind outflow.







\acknowledgments
We acknowledge helpful discussions with S.\ E.\ Guidoni and P.\ F.\ Wyper. The work of V.M.U.\ was supported by NASA grant NNG11PL10A 670.036 to CUA / IACS. J.T.K.\ and C.R.D.\ were supported by a NASA LWS grant to investigate solar coronal jets and their heliospheric consequences.  The numerical simulations were performed under a NASA High-End Computing award to C.R.D.\ on Discover, the NASA Center for Climate Simulation computing system.

\bibliographystyle{apj} 
\bibliography{apj_2015}

\end{document}